\documentclass[print,linenumbers]{aa}
\usepackage{graphicx}
\usepackage{multirow}

\usepackage[varg]{txfonts}
\usepackage{cases}
\usepackage{natbib}
\usepackage{underscore}
\usepackage{hyperref}

\usepackage{amsmath}
\usepackage{threeparttable}
\usepackage{placeins}

\usepackage[switch]{lineno}

\titlerunning{Filament Deflection via Reconnection at an X-Point}

\begin{document} 

   \title{Deflection of a Filament Eruption with Three Parallel Flare Ribbons via Reconnection at an X-Point}

   \subtitle{}

   \author{Xiaomeng Zhang\inst{1}
          \and
          Jinhan Guo\inst{1, 2}
          \and
          Yang Guo\inst{1}
          \and
          Mingde Ding\inst{1}
          \and
          Brigitte Schmieder\inst{2}
          }

    \institute{School of Astronomy and Space Science and Key Laboratory of Modern Astronomy and Astrophysics, Nanjing University, Nanjing 210023, China \\
      	\email{guoyang@nju.edu.cn}
 	    \and 
 	    Centre for Mathematical Plasma Astrophysics, Department of Mathematics, KU Leuven, Celestijnenlaan 200B, B-3001 Leuven, Belgium}

   \date{}
 
  \abstract
   {On 2024 May 6, Active Region 13663 produced an X4.5-class flare associated with a filament eruption that exhibited remarkable rotation and deflection dynamics.}
   {This study aims to investigate two key aspects of this event: the formation mechanisms of the complex flare ribbon structures and the physical drivers behind the observed filament deflection.}
   {We conduct a data-constrained magnetohydrodynamic simulation using the zero-beta approximation to reconstruct the filament's evolution. Through detailed analysis of quasi-separatrix layers (QSLs) and their comparison with observed flare ribbons, we establish crucial connections between magnetic topology and flare morphology.}
   {First, our simulation successfully reproduces key observational features of the eruption. Then, we connect the flare ribbon morphology with calculated QSLs. Finally, we find filament deflection resulting from localized reconnection at the X-point, as evidenced by Lorentz force decomposition.} 
   {We demonstrate that reconnection above two current channels of opposite helicity governs the eruption dynamics, with magnetic pressure gradients driving flux rope deflection while magnetic tension force simultaneously restraining arcade ascent. The event features a "sandwich" magnetic configuration including double parallel polarity inversion lines with strong shear component. We suggest that this particular configuration could serve as a plausible formation mechanism for the observed parallel three-ribbon structure. In addition, the evolution of QSLs and flare ribbons provides clear evidence of reconnection between two flux ropes.}


\maketitle

\section{Introduction}\label{introduction} 

Solar eruptions, such as filament eruption and coronal mass ejections (CMEs), usually exhibit propagation deviating significantly from radial trajectories. These events challenge traditional eruption models and always suggest complex interactions between magnetic flux ropes, overlying field structures, and external driving forces. Understanding the mechanisms behind non-radial eruptions, such as rotation and deflection, is crucial for improving space weather predictions.

Extensive researches have investigated the mechanisms underlying filament rotation and deflection during solar eruptions. Regarding rotational dynamics, \citet{Torok2005} demonstrated that eruptions triggered by kink instability invariably exhibit rotational motion. \citet{Isnberg2007} established that flux ropes undergo rotation due to Lorentz forces exerted by external toroidal fields. Through comprehensive parameter studies, \citet{Kliem2012} systematically quantified flux-rope dynamics under varying external toroidal fields. Inspired by their work, \citet{Zhang2025} identified distinct interaction modes between toroidal fields and flux ropes. They also suggested that slipping reconnection can enhance rotational motion. \citet{Williams2005} and \citet{Guo2023} further showed that magnetic reconnection between flux ropes and external fields can significantly modify the orientation of the flux rope axis. Additionally, \citet{Zhang2024} demonstrated that interactions in double flux-rope systems can mutually drive rotational behavior.

As for deflection, the coronal magnetic configuration always plays a pivotal role in determining CME trajectories, with flux ropes being repelled by regions of high magnetic pressure (e.g., strong active regions and coronal holes) while being attracted to areas of low magnetic pressure (e.g., streamer belts and the heliospheric current sheet). This fundamental relationship has inspired studies to calculate distributions of magnetic energy density for deflection predictions \citep{Shen2011, Gui2011, Liu2018}. Besides magnetic topology, multiple factors influence CME deflection. For example,  \citet{Kay2015} demonstrated that lower-mass, slower CMEs exhibit greater susceptibility to deflection. \citet{Wangym2014} revealed that velocity differentials between interplanetary CMEs (ICMEs) and Parker-spiral-structured solar wind can induce significant trajectory deflection. Frequent collisions between ICMEs with different velocities can also result in mutual deflection \citep{Lugaz2012}. For comprehensive discussions of CME deflection mechanisms, we refer readers to reviews by \citet{Manchester2017} and \citet{Cecere2023}.

Observational studies have documented numerous cases of filament deflection during eruptions. \citet{Jiang2009} observed a filament situated at the inner boundary of a streamer structure, which subsequently underwent lateral ejection along the streamer's outline. In an analysis of sympathetic partial eruptions, \citet{Yang2023} documented two filaments with opposite magnetic helicity that erupted sequentially, with their mutual interaction inducing significant deflection. \citet{Kay2017} systematically tracked deflections in seven CMEs originating from Active Region 11158 during 2011 February 13--16, revealing that CMEs from the same active region can follow markedly divergent propagation paths. \citet{Qiu2025} showed that non-radial filament eruptions preferentially propagate toward regions of weaker overlying poloidal magnetic field, highlighting the role of asymmetric strapping forces in shaping filament trajectories. Recognizing that both the magnetic pressure gradient and magnetic field configuration influence eruption trajectories, \citet{Sahade2025} conducted a comparative study of eight CMEs, demonstrating that topological path provides superior prediction of low-coronal deflection patterns compared to gradient path.

Most observational studies of filament deflection rely on multi-wavelength evolution analysis and nonlinear force-free field (NLFFF) extrapolations. However, time-dependent simulations of such deflections remain scarce, which is a gap also highlighted by \citet{Cecere2023}. Furthermore, while many deflection studies examine phenomena at scales of several solar radii, few focus on deflection dynamics within active regions. To address these limitations, we perform a data-constrained magnetohydrodynamic (MHD) simulation to reproduce an observed filament eruption exhibiting clear deflection within an active region. Section \ref{method} details the observational characteristics and numerical setup of our simulation. In Section \ref{result}, we compare simulation outputs with observations and analyze the physical mechanisms driving the deflection. Finally, Section \ref{discussion} summarizes our findings and discusses the key properties of the observed magnetic field configuration.

\section{Observation and Numerical setup}\label{method} 

The filament eruption occurred at solar coordinates (N25, W35) in the NOAA Active Region (AR) 13663. This event is associated with an X4.5-class solar flare, which commenced at 05:38 UT, reached its peak intensity at 06:35 UT, and ended at 06:52 UT on 2024 May 6. This eruption was captured by the Atmospheric Imaging Assembly \citep[AIA;][]{Lemen2012} aboard the Solar Dynamics Observatory \citep[SDO;][]{Pesnell2012}. Before the major flare, two localized britenings occurred in the southern part of the active region. Throughout the eruption, the filament exhibited pronounced non-radial dynamics, characterized by a counterclockwise rotational motion and a deflection towards the southwest. Concurrently, the flare ribbons underwent a complex evolutionary sequence, transitioning from initial point-like brightenings to the subsequent separation of multiple ribbon structures. 

\begin{figure*}[ht!]
\centering
\includegraphics[scale=0.5]{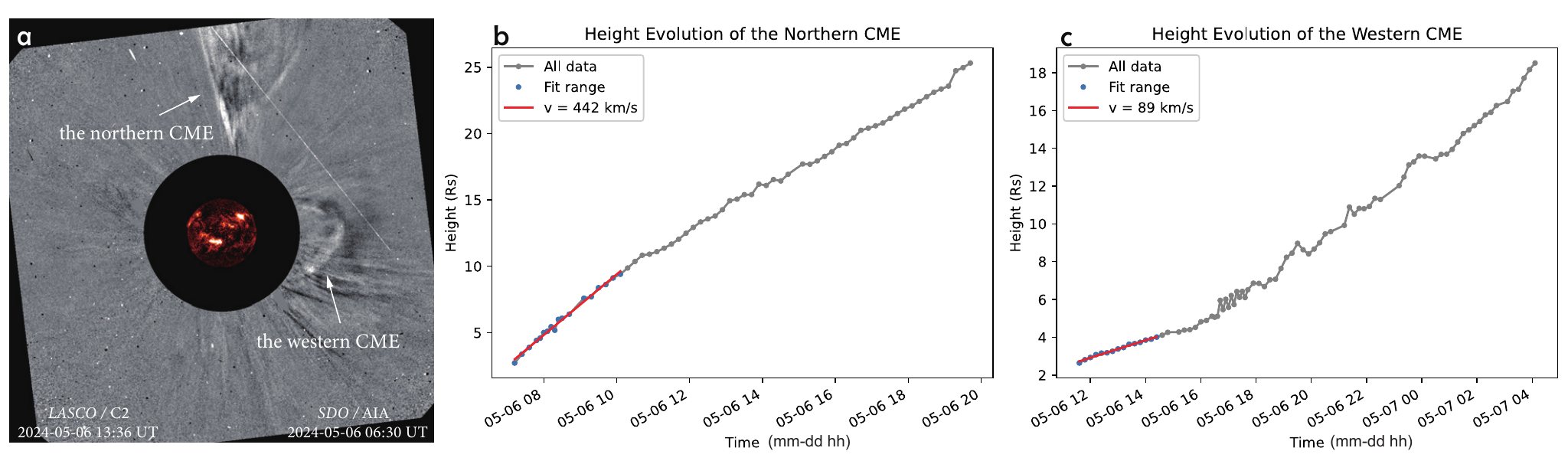}
\caption{(a) Two CMEs are visible in the \emph{SOHO}/LASCO C2 field of view at 13:36 UT on 2024 May 6: an nCME and a wCME. The inset shows the \emph{SDO}/AIA 304~{\AA} view at 06:30 UT on the same day. (b) Height evolution of the nCME. Blue dots indicate the data points used to derive the velocity, and the red line shows the linear fit with a velocity of 442~km~s$^{-1}$. (c) Same as panel (b), but for the wCME, with a fitted velocity of 89~km~s$^{-1}$.}
\label{fig1}
\end{figure*}

Subsequently, the filament underwent a successful plasma ejection, manifesting as a CME observable in the field of view of the Large Angle and Spectrometric Coronagraph \citep[LASCO;][]{Brueckner1995} C2 aboard the Solar and Heliospheric Observatory \citep[SOHO;][]{Domingo1995} several hours after the eruption. Interestingly, two CMEs appear in the LASCO C2 field of view following the filament eruption: a northern CME (nCME) and a western CME (wCME; Figure~\ref{fig1}a). Unfortunately, due to the lack of observations between 1.3 and 2.2~$R_{\odot}$, we cannot establish a definitive connection between the CMEs and the filament eruption. However, by comparing their propagation directions and estimating the CME onset times through backward extrapolation of their measured speeds, we suggest that both CMEs originate from AR 13663, and that the wCME is associated with the filament eruption. A more detailed analysis will be presented in Section~\ref{sub:comparison}.

In this study, we concentrate on elucidating the underlying mechanisms responsible for the non-radial motion of the filament and the formation of flare ribbons. To uncover these mechanisms, a data-based simulation is essential to accurately reproduce and comprehensively interpret the observational phenomena.

Similar to previous works \citep{Guo2019,Zhong2021,Zhang2024}, we conduct a data-constrained simulation with zero-$\beta$ assumption, and the governing MHD equations are as follows: 

\begin{eqnarray} 
&& \frac{\partial \rho}{\partial t} +\nabla \cdot(\rho \boldsymbol{v})=0,\label{eq1}\\
 && \frac{\partial (\rho \boldsymbol{v})}{\partial t}+\nabla \cdot(\rho \boldsymbol{vv}-\boldsymbol{BB})+\nabla (\frac{\boldsymbol{B}^2}{2})=0,\label{eq2}\\
 && \frac{\partial \boldsymbol{B}}{\partial t} + \nabla \cdot(\boldsymbol{vB-Bv})=0,\label{eq3}
\end{eqnarray}
where $\rho$, $\boldsymbol{v}$, $\boldsymbol{B}$ are the density, velocity, and magnetic field, respectively. All physical quantities are normalized by typical coronal values: $\rho_{0}=2.3\times 10^{-15}$ $\rm g\ cm^{-3}$, $v_{0}=1.2\times 10^{7}$ $\rm cm\ s^{-1}$, $t_{0}=85.9$ s, and the normalized length $L_{0}=1.0\times 10^{9}$ cm. 

\begin{figure*}[ht!]
\centering
\includegraphics[scale=0.8]{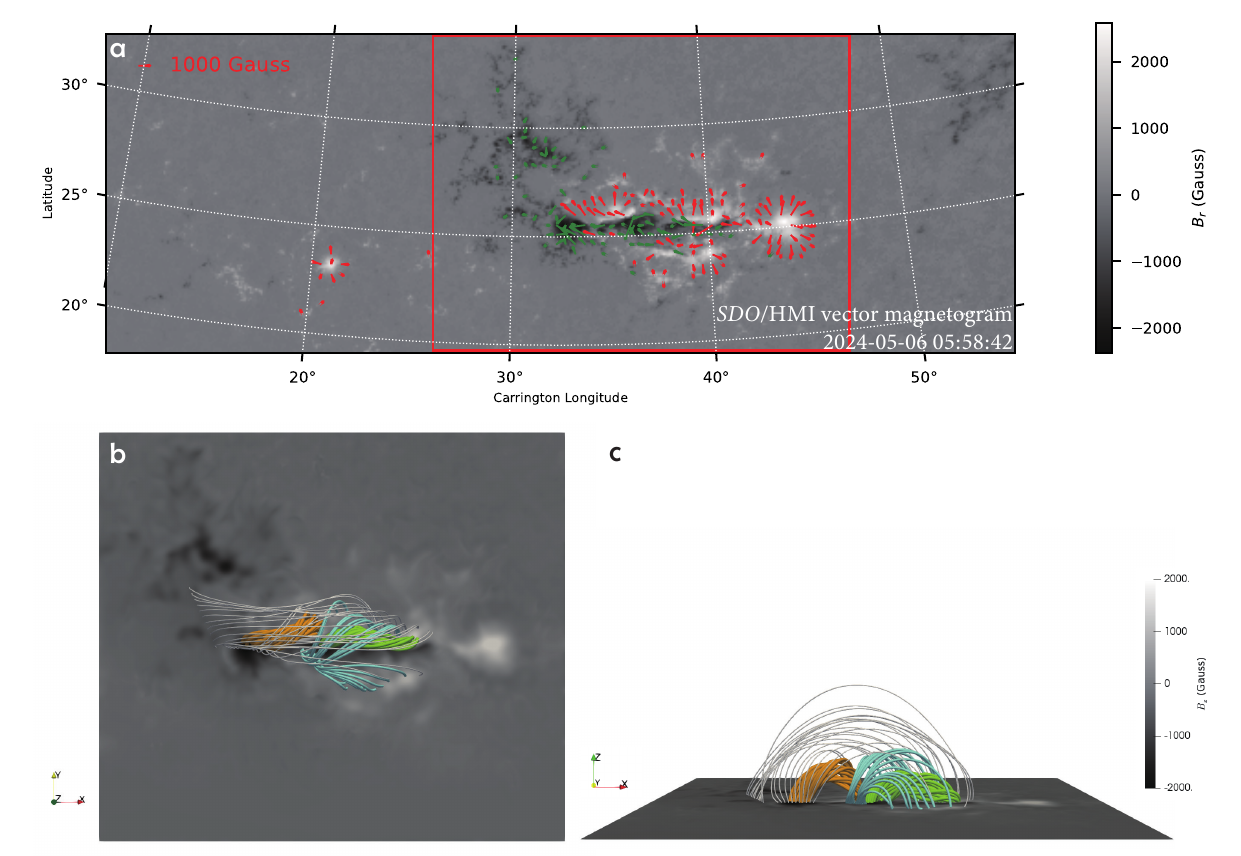}
\caption{(a) \emph{SDO}/HMI vector magnetogram in the CEA coordinate system at 05:58 UT on 2024 May 6, where the red (green) arrows represent the directions of the horizontal field in positive (negative) polarities. The region enclosed by the red rectangle is selected and processed as the bottom boundary magnetic field for the extrapolation. (b) Top view of magnetic configuration in an NLFFF model. The western (eastern) flux rope is colored in green (orange). The lower (higher) arcades are colored in cyan (gray). (c) Side view of the same configuration as in panel (b).}
\label{figure1}
\end{figure*}

The simulation begins with an NLFFF extrapolation, whose bottom vector magnetic field is observed at 05:58 UT on 2024 May 6 provided by the Helioseismic and Magnetic Imager \citep[HMI;][]{Scherrer2012, Schou2012} on board \emph{SDO}. The vector magnetic field data, derived from the Spaceweather HMI Active Region Patch (SHARP) series, has been processed by removing the $180^\circ$ ambiguity and correcting projection effects. For the small field of view around the center of AR 13363, the spherical SHARP magnetogram is approximated by a local tangent plane, and the Cylindrical Equal-Area (CEA) components $(B_r, B_t, B_p)$ are transformed into Cartesian components $(B_z, -B_y, B_x)$ with negligible distortion. The processed data are visually presented in Figure \ref{figure1}a. After the removal of the net Lorentz force and torque, the region in the red box is used for extrapolation. After relaxation with the magnetofrictional (MF) method \citep{Guo2016a, Guo2016b}, the force-free metric is $\sigma_J=0.35$ and the divergence-free metric is $|f_{i}|\rangle=1.53 \times 10^{-5}$. The $\sigma_J$ may not be very low, as this active region exhibits a strong shear component and is unlikely to relax into a fully force-free state. The magnetic field configuration derived from the NLFFF extrapolation is presented in Figures \ref{figure1}b and \ref{figure1}c, displaying top and side views, respectively. The configuration reveals two distinct flux ropes situated beneath lower and higher magnetic arcades. Notably, the right flux rope exhibits a strong spatial correspondence with the observed filament, as shown in Figure~\ref{fig2}a. In contrast, the left flux rope does not have a corresponding filament and can only be identified through the associated flare ribbons. The left flux rope might be a strong candidate for driving the nCME shown in Figure~\ref{fig1}a. This possibility will be discussed further in Section~\ref{sub:comparison}.

The initial density distribution is the same as \citet{Guo2019}, which is given by a stratified atmospheric model, and the initial velocity is set to zero everywhere. The MHD equations are solved with the Message Passing Interface Adaptive Mesh Refinement Versatile Advection Code \citep[MPI-AMRVAC\footnote{http://amrvac.org},][]{Xia2018, AMRVAC2023} in the local Cartesian coordinate system. The simulation domain is $[x_{\min}, x_{\max}] \times \allowbreak [y_{\min}, y_{\max}] \times \allowbreak [z_{\min}, z_{\max}] = \allowbreak [-117, 117] \times \allowbreak [-102, 102] \times \allowbreak [1, 147]$ Mm, which is resolved by ${320\times280\times200}$ cells. Specifically, we employ a uniform grid with a resolution of 700 km for the computational domain. Similar to \citet{Guoy2021}, the bottom boundary magnetic field remains fixed in the data-constrained simulation, while a second-order zero-gradient extrapolation is applied to the remaining five boundaries. The velocity at the bottom boundary is set to zero, with open boundary conditions implemented for the other boundaries. Additionally, a zero-gradient extrapolation is adopted for the density across all boundaries. The time units in Figures \ref{figure2}, \ref{figure3}, \ref{figure4}, and \ref{figure5}, as well as in the subsequent text, are all 0.7 $t_0$, which corresponds to approximately 1 minute.

In Section \ref{sub:deflection}, we calculate the Lorentz force and its components using the following equation:
\begin{equation}
  \mathbf{F} = \frac{1}{\mu_0}\ (\mathbf{B}\ \cdot\ \nabla)\ \mathbf{B}\ -\ \nabla\ (\frac{B^2}{2\mu_0})
  \label{eq4}
\end{equation}
where vacuum permeability $\mu_0=4\pi \times 10^{-9}\ \rm H \cdot cm^{-1}$. The first term on the right-hand side of the equation represents the magnetic tension force, while the second term corresponds to the magnetic pressure. To analyze the actual contributions of these force components to magnetic field dynamics, we decompose them into their perpendicular projections relative to the local magnetic field direction. The resulting perpendicular components, the magnetic tension force ($\mathbf{T}$) and the magnetic pressure force ($\mathbf{P}$), are defined as follows:
\begin{eqnarray}
&& \mathbf{T} = \frac{B^2}{\mu_0}(\mathbf{b}\ \cdot\ \nabla)\mathbf{b},\label{eq5}\\
&& \mathbf{P} = (\mathbf{bb-I}) \cdot \nabla\ (\frac{B^2}{2\mu_0}),\label{eq6}
\end{eqnarray}
where $\mathbf{b} = \mathbf{B}/|B|$ defines the unit vector aligned with the local magnetic field direction. $\mathbf{I}$ is the identity tensor in Cartesian coordinate system.

\section{Result}\label{result}
\subsection{Comparisons between Simulation and Observations} \label{sub:comparison}
\begin{figure*}[ht!]
\centering
\includegraphics[scale=0.5]{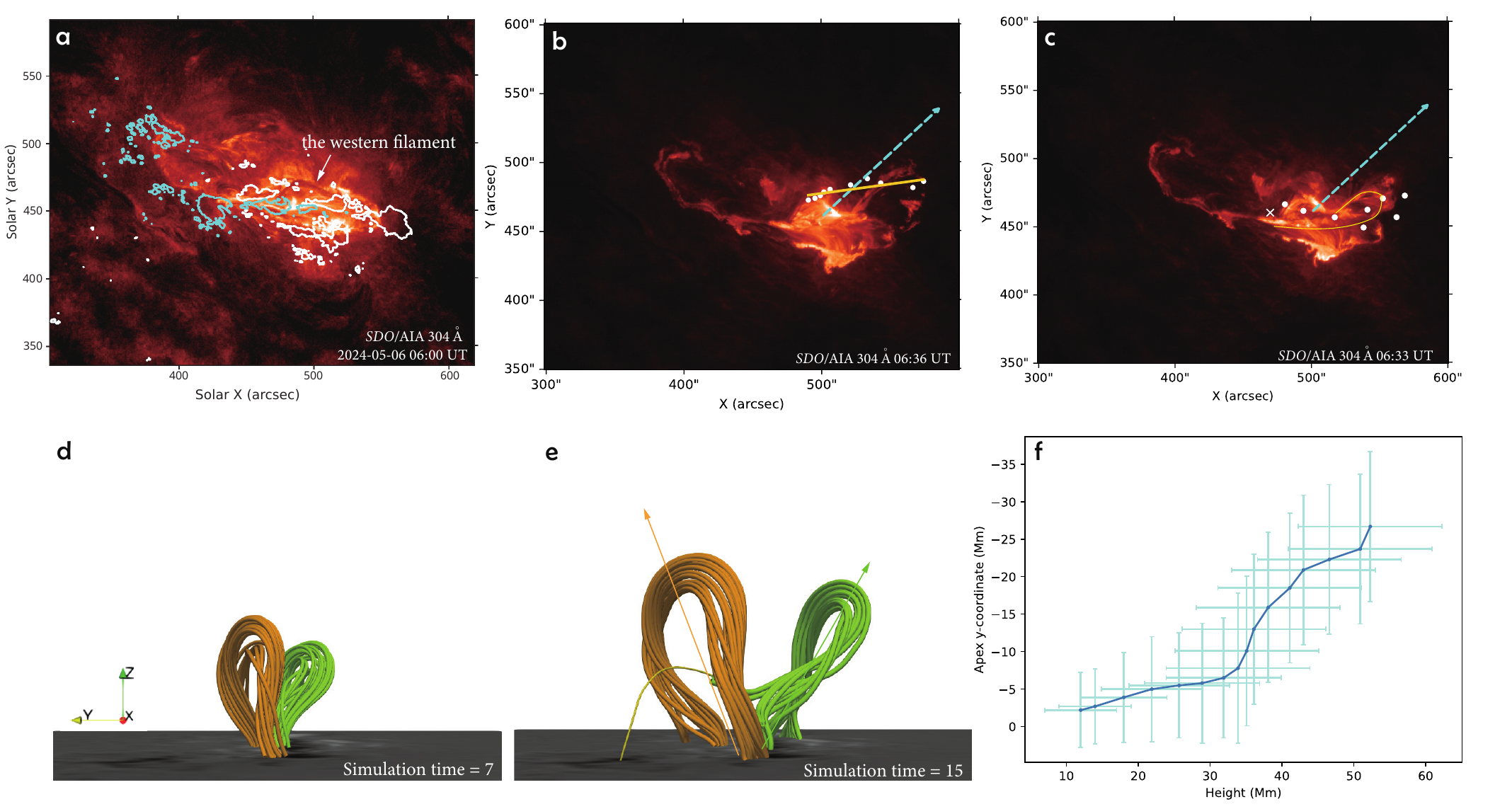}
\caption{(a) \emph{SDO}/AIA 304 {\AA} image at 06:00 UT, which is covered by the white (cyan) contour of positive (negative) magnetic field in the line of sight observed by \emph{SDO}/HMI. The white arrow indicates the position of the western filament. (b) Comparison between the filament eruption direction and the radial direction. The white dots mark the locations of the filament apex in the AIA plane of sky at different observation times. The yellow line represents a linear fit to these points. The cyan dashed arrow indicates the radial direction projected onto the AIA view. The background shows the AIA image observed at 06:36~UT. (c) Comparison between the shapes of the simulated flux rope and the observed filament. The white dots show the projected shape of the flux rope in the AIA view at simulation time $t = 15$, while the white cross marks the eastern footpoint of the flux rope. The yellow outline highlights the filament observed at 06:33 UT. The cyan dashed line is the same as that shown in panel (b). (d) Two flux ropes at simulation time $t = 7$. The orange (green) flux rope represents the eFR (wFR). (e) Same as panel (d), but at simulation time $t = 15$. Two arrows indicate the approximate propagation directions of the flux ropes. Note that several field lines of the green flux rope are not shown completely in order to keep the image clear. A thin yellow line indicates the full shape of the partially omitted green field lines. (f) Variation of the $y$-coordinate of the flux-rope apex as a function of height in the simulation. The light-blue plus symbols at each point indicate the approximate flux-rope radius at the corresponding time.}
\label{fig2}
\end{figure*}

Before performing a detailed comparison between the simulation and the observations, we first present the deflection of the filament inferred from the observations and that of the flux rope obtained from the simulation. Figure~\ref{fig2}b shows the observed deflection of the filament. The cyan dashed line represents the radial direction projected onto the AIA plane of the sky. It is evident that the filament erupts along a direction that deviates from the radial direction by approximately $40^\circ$ in projection. In the simulation, the two extrapolated flux ropes deflect in opposite directions: the eastern flux rope (eFR) deflects toward the north, while the western flux rope (wFR) deflects toward the south (Figures~\ref{fig2}d and e). We quantify the wFR's deflection in the simulation by tracking the position of its apex in the $y$-direction, as shown in Figure~\ref{fig2}f. The apex of the wFR undergoses a continuous southward motion with a displacement of about 25~Mm in the $y$-direction over the course of the simulation. To verify the plausibility of this southward deflection, we back-project the wFR at simulation time $t=15$ onto the AIA plane of the sky and compare it with the observed filament configuration at 06:33~UT. In Figure~\ref{fig2}c, a small offset of a few arcseconds is present between the two outlines, suggesting that the simulated flux rope is either located at a slightly lower height or experiences a larger deflection in the $y$-direction compared to the observed filament. Given that the flux-rope radius is approximately 10~Mm at this time, this displacement is within acceptable uncertainty and does not affect the validity of the simulation for investigating the deflection mechanism. It should be noted that although the magnetogram at 05:58 UT serves as the bottom boundary for the extrapolation, the resulting flux rope, while in equilibrium, remains unstable and susceptible to eruption under minor numerical perturbations. Consequently, the observational data corresponding to the initial stage of the filament eruption (06:18 UT) are used for comparison with the simulated flux rope at its initial state.

Using the directions of the two flux ropes and the velocities of the observed CMEs, we evaluate the correspondence between the simulated flux ropes and the CMEs. First, we assess the temporal relationship by estimating the onset times of the two CMEs through a linear backward extrapolation of their velocities. In Figures~\ref{fig1}b and c, the early propagation of the two CMEs in the LASCO C2 field of view is fitted linearly, yielding initial velocities of approximately 442~km~s$^{-1}$ for the nCME and 89~km~s$^{-1}$ for the wCME. By extrapolating the fitted velocity profiles back to the solar surface, the CME onset times are estimated to be around 06:01~UT and 05:52~UT, respectively. If the location of the source active region (heliographic coordinates N25, W35) is taken into account, corresponding to a projected height of approximately 0.6~$R_\odot$, the estimated eruption times become about 06:17~UT and 07:01~UT. The filament eruption occurs at 06:18~UT, which is very close to the time interval of 06:01–06:17~UT and lies within the broader uncertainty range of 05:55–07:10~UT. Therefore, both CMEs are temporally consistent with the eruption in AR 13663. 

Next, we examine the spatial relationship. In the AIA plane-of-sky view, the wCME propagates approximately along the extension of the filament eruption direction. Although the eFR does not have a corresponding filament, its northward deflection in the simulation suggests that it may be the source of the nCME. In addition, we inspect the observations from the Full Sun Imager (FSI) of the Extreme Ultraviolet Imager (EUI) on board \textit{Solar Orbiter} \citep{solo}. With a separation angle of about 150$^\circ$ from \textit{SDO}, this viewpoint provides a near-backside view of the Sun. No strong eruption is detected on the far side during this period. Taken together, these results suggest that AR~13663 may have produced two CMEs nearly simultaneously: a strong and fast nCME and a weaker and slower wCME. Because the eFR does not have a clear filament counterpart, we do not focus on it further and instead concentrate on the deflection of the wFR, which corresponds to the observed filament eruption.

\begin{figure*}[htbp]
\centering
\includegraphics[scale=0.5]{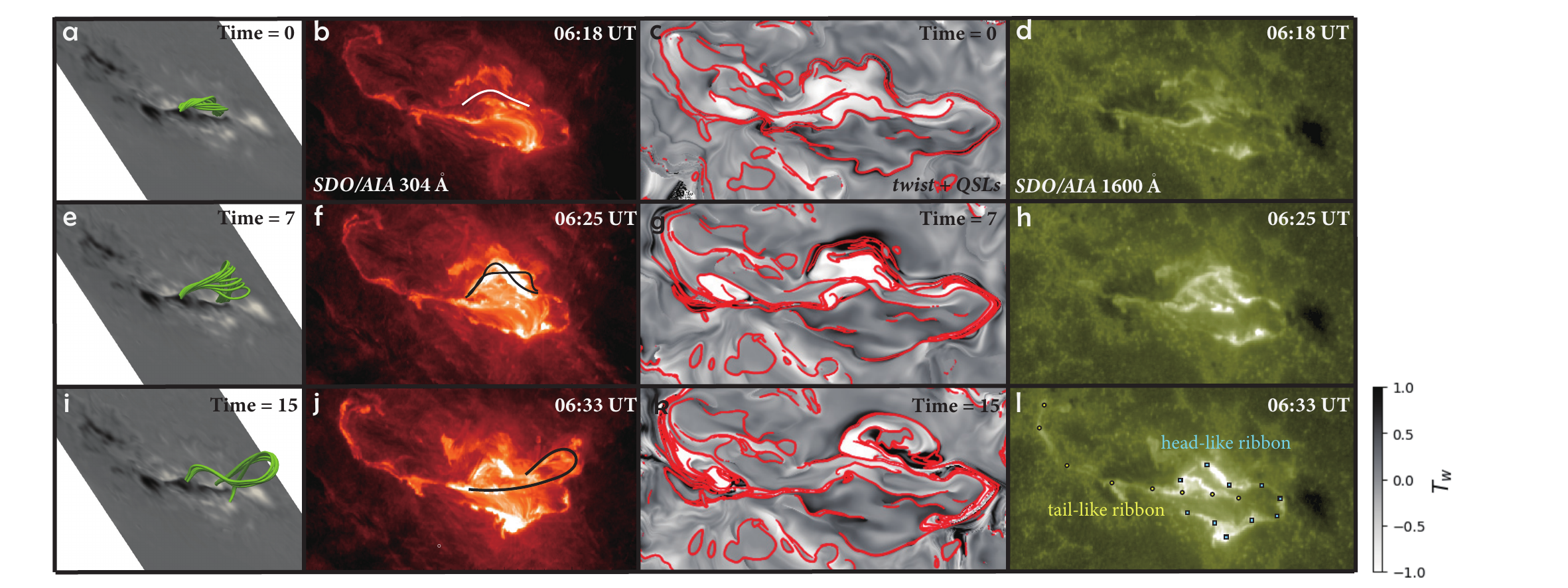}
\caption{Panels (a), (e), and (i) depict the evolution of the wFR from an observational perspective at simulation times 0, 7, and 15, respectively. Panels (b), (f), and (j) display \emph{SDO}/AIA 304 \AA \ images capturing the filament eruption at 06:18, 06:25, and 06:33 UT, respectively. The white and black lines delineate the trajectory of the filament at the respective times. Panels (c), (g), and (k) present the photospheric twist distribution overlaid with red contours of QSLs ($log\ Q = 7$) at the same simulation times as panels (a), (e), and (i), respectively. Panels (d), (h), and (l) exhibit \emph{SDO}/AIA 1600 \AA \ images illustrating the evolution of flare ribbons at the same observation times as panels (b), (f), and (j), respectively. In panel (l), the spatial distributions of tail-like ribbon and a head-like ribbon are demarcated by yellow circles and blue squares, respectively.}
\label{figure2}
\end{figure*}

After presenting the deflection, we proceed to a detailed comparison between the simulation and the observations. Given the different observational perspectives of the simulation and observation, we reorient the simulation box to align with the observational view, enabling a direct comparison between the flux-rope evolution and the filament eruption. In our simulation, the wFR corresponds to the observed filament. To comprehensively illustrate the evolutionary process, we select three representative times characterized by distinct morphological features of the flux rope and compare them with the corresponding observational timestamps. The comparison is shown in Figures \ref{figure2}a, b, e, f, i, and j (first and second columns), where the filament path is outlined to present its shape clearly. The simulation results demonstrate that the flux rope undergoes a distinct counterclockwise rotation accompanied by a southward deflection. These kinematic features show excellent agreement with the observed filament evolution, both in terms of morphological changes and rising velocity characteristics.

Subsequently, we calculate the squashing factor ($Q$), a quantitative measure that characterizes the topological complexity of the magnetic field and indicates potential sites for magnetic reconnection. The computational methodology follows the approach proposed by \citet{Scott2017}, implemented through an open-source numerical routine \citep{Zhang2022}. In parallel, we derive the magnetic twist distribution using the same computational framework. In Figures \ref{figure2}c, g and k, we superimpose the photospheric twist distribution with contours of Quasi-Separatrix Layers (QSLs), regions characterized by significantly enhanced $Q$ values ($Q \gg 2$) that typically mark preferential locations for intense magnetic reconnection activity. The twist distribution is fundamentally partitioned into two distinct regions by the QSLs: a positive twist region bounded by the central and northern QSLs, and a negative twist region enclosed by the central and southern QSLs. Notably, the central QSLs exhibit spatial coincidence with the strip shaped negative magnetic field in this region, indicating that this negative magnetic field serves as a topological boundary separating domains of opposite magnetic helicity. 

\begin{figure*}[ht]
\centering
\hspace*{-1cm} 
\includegraphics[scale=0.52]{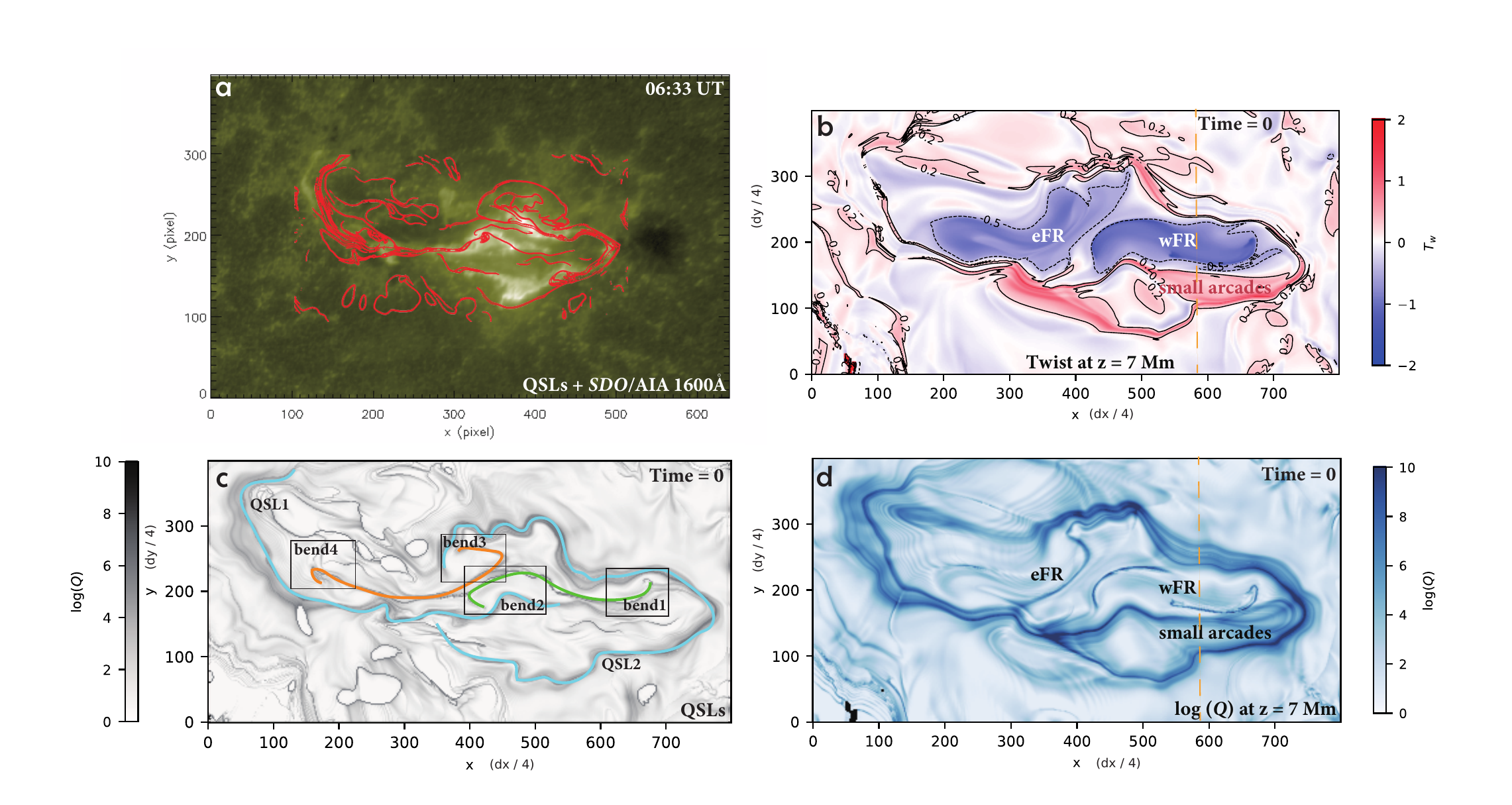}
\caption{(a) \emph{SDO}/AIA 1600 \AA \ image at 06:33 UT overlaid with QSLs in red contours at simulation time = 15. (b) Twist distribution at z = 7 Mm at simulation time = 0. Positions of eFR, wFR and small arcades are marked. The black dashed and solid lines represent the twist = -0.5 and twist = 0.2 contours, respectively. The orange dash line shows the slice location in Figure \ref{figure4}a and b. (c) log ($Q$) distribution at z = 0 at simulation time = 0, with high-Q regions highlighted in color. Blue lines mark the locations of short tail-like ribbon and head-like ribbon, while orange and green lines mark two S-shaped riboons, respectively. Black rectangles denote the positions of four bends, which are corresponding to hooks observed in flare ribbons. (d) log ($Q$) distribution at z = 7 Mm at simulation time = 0 with same markers as in panel (b). The pixel size in panels (b)–(d) is dx/4 = dy/4 = 175 km, where dx = dy = 700 km is the spatial resolution of the simulation.}
\label{figure3}
\end{figure*}

To further testify the validity of our simulation, similar to previous works \citep{Guo2019,Zhong2021,Zhang2024}, we compare the QSL evolution with observed flare ribbons. Figures \ref{figure2}d, h, and l illustrate the evolution of the flare during the eruption. The flare ribbons exhibit a distinct tadpole-like morphology, resembling that reported in \citet{Vemareddy2014}. Two primary ribbons can be identified: the “tail” of the tadpole, corresponding to an elongated ribbon extending from the top-left corner toward the center of the field of view, and the “head” of the tadpole, characterized by a compact structure with its opening facing left. Although the overall ribbon morphology is similar to that observed in \citet{Vemareddy2014}, the underlying magnetic configuration and the eruptive outcome differ between the two events. These differences may be related to the stronger magnetic shear and larger amount of free magnetic energy present in our case. During the eruption, each segment of the two ribbons brightens sequentially, gradually changes its shape, and eventually assumes a configuration that closely resembles the distribution of QSLs. A more distinct comparison of flare ribbons and QSLs is shown in Figure \ref{figure3}a. Similar to \citet{Zhao2016}, we overlaid the observed flare ribbons with QSLs to compare their shapes. Since the flare ribbons and QSLs are represented in different coordinate systems, we remapped the \emph{SDO}/AIA 1600 \AA \ observation at 06:33 UT to the CEA coordinate system and overlaid the QSL contour at simulation time = 15 onto it. The primary features of the QSLs exhibit a strong correspondence with the flare ribbons.

Then, utilizing the simulation data, we establish the correspondence between the flare ribbon structure and the magnetic field configuration in Figure \ref{figure3}c. To clearly illustrate the topological properties, we employ the initial QSL distribution as the background due to its well-defined shape and partitioning. Specifically, the tail-like ribbon is not a single cohesive structure but rather consists of three distinct segments: a shorter tail-like ribbon (outlined in blue and labeled as "QSL1") and two S-shaped ribbons (outlined in orange and green). The two S-shaped ribbons delineate the locations of two flux ropes, with their four bends representing the footpoints. These bends correspond spatially to the observed hooks in the flare ribbons, which are the characteristic signatures of flux rope footpoints \citep{Titov2007,Pariat&Démoulin2012,Aulanier2025}. The head-like ribbon in Figure \ref{figure2}l spatially coincides with the quasi-separatrix layer QSL2 (delineated in blue in Figure \ref{figure3}c). Furthermore, QSL1 and QSL2 are associated with the negative and positive footpoints of a set of coronal arcades in our simulation, respectively. Representative examples of these arcades include the lower (cyan) and higher (gray) structures illustrated in Figures \ref{figure1}d and e. This spatial correspondence suggests that the flare ribbons co-located with QSL1 and QSL2 were formed by magnetic reconnection driven by the stretching of overlying arcades during the ascent of the flux rope. To more clearly distinguish the two flux ropes, we also plot the distributions of twist and log($Q$) on a higher plane at z = 7 Mm. As shown in Figure \ref{figure3}b, regions of negative twist bounded by the twist = –0.5 contour clearly delineate the two flux ropes. The corresponding log($Q$) distribution on the same plane (Figure \ref{figure3}d) reveals current channel structures consistent with that identified by the twist distribution.

It is interesting to analyze the kinematic behavior in the bend/hook structures, particularly bends 2 and 3 and their corresponding hooks. Initially spatially separated and associated with different flux ropes (Figure \ref{figure2}c), these bends exhibited convergent motion upon eruption onset. By 06:25 UT, both observational flare ribbons and QSL distributions (time = 7) demonstrated the formation of a transient closed topology (Figures \ref{figure2}g and h). Subsequent evolution showed progressive retraction of this structure, culminating in complete spatial coincidence of the bend2 and bend3 boundaries by 06:33 UT, as confirmed by both flare ribbons and QSL analysis (Figures \ref{figure2}k and l). This footpoint coalescence suggests a novel connectivity between the flux ropes, analogous to yet distinct from classical tether-cutting reconnection scenarios \citep{Zou2019}, as it involves interaction dynamics different from typical sheared arcade configurations.

\subsection{Filament deflection} \label{sub:deflection}

\begin{figure*}[ht]
\centering
\includegraphics[scale=0.6]{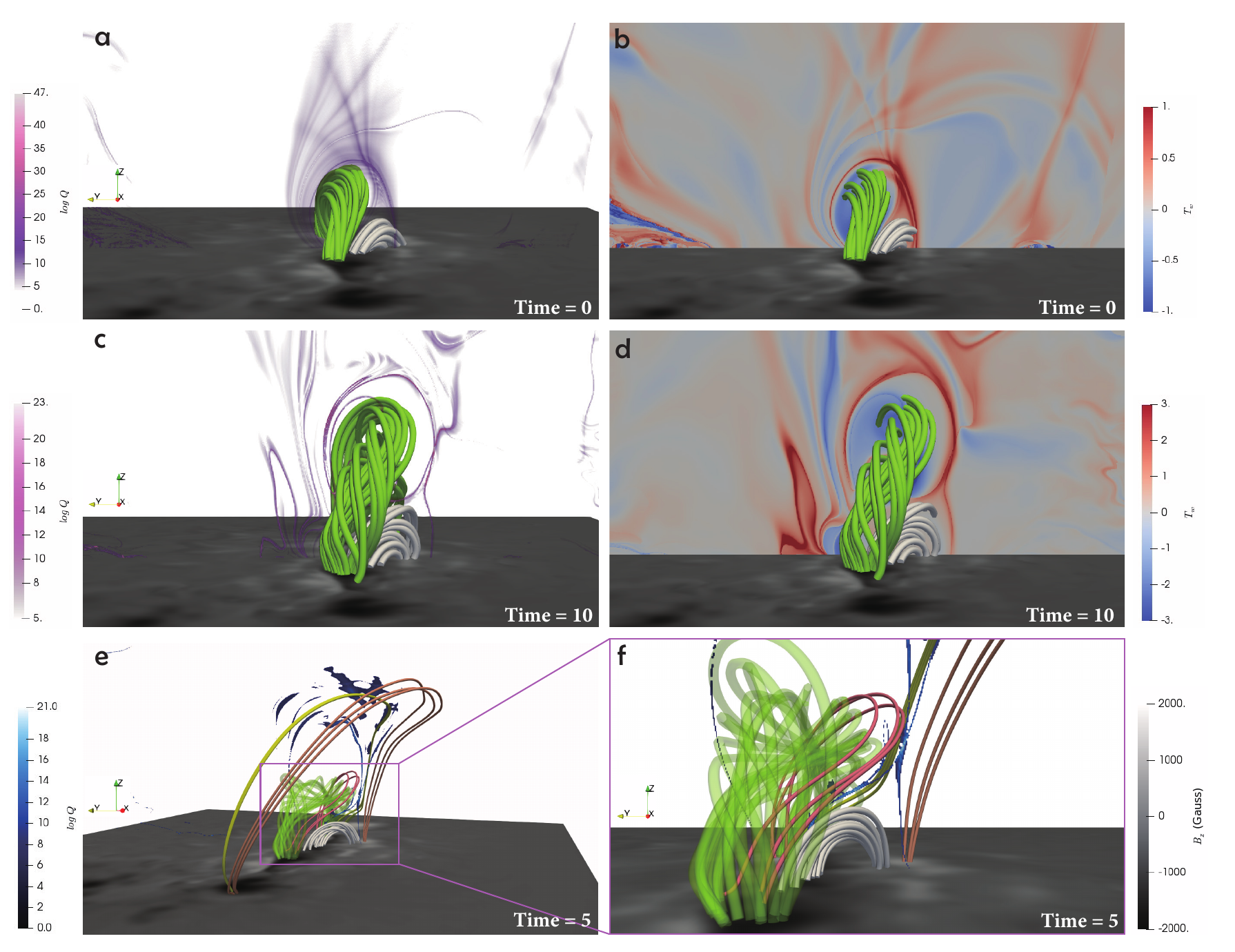}
\caption{(a) Magnetic configuration of two current channels at simulation time = 0. The wFR is colored in green and the small arcade is colored in gray. The semi-transparent slice shows the QSL distribution. (b) Magnetic configuration is the same as panel (a), but the slice shows twist distribution. The slice locates at $x=31.4$ Mm. (c) Same as panel (a), but at simulation time = 10. (d) Same as panel (b), but at simulation time = 10. The slices shown in panels (c) and (d) are taken at $x=20.0$ Mm. (e) Magnetic configuration around the X-point. The dark blue patches represent QSL contours. The semi-transparent green flux rope corresponds to wFR, with several of its peripheral field lines highlighted in red. The small arcade and several field lines formed through reconnection are depicted in gray. Two sets of external field lines are shown in orange and yellow, respectively. (d) a zoomed-in view of the configuration shown in panel (c).}
\label{figure4}
\end{figure*}

To investigate the cause of the filament deflection, we select a vertical slice perpendicular to the initial wFR, which corresponds to the observed filament, and analyze the QSL distribution on this plane. As shown in Figures \ref{figure4}a and c, two current channels traverse two distinct regions separated by high-Q layers. In addition to wFR (highlighted in green), a small system of arcades (colored in gray) passes through another region enveloped by QSLs. Note that these small arcades, located in the southern region of the wFR, should not be confused with the eFR. Then, we calculate the twist distribution on the same slices in Figures \ref{figure4}b and d and find that the two current channels exhibit opposite twists: wFR possesses negative helicity, while the arcades display positive helicity. This difference in helicity is also evident in Figure \ref{figure3}b, which illustrates the relative positions of wFR and the small arcades, along with their twist distribution on the z = 7 Mm plane. Due to their distinct topological characteristics, high-Q regions form around these current channels. Examining the field lines surrounding the current channels, the QSLs manifest as an X-shaped structure, indicating that magnetic reconnection is likely to occur at this X-point. It is important to note that this X-point is not a null point, as an inward component of the magnetic field is present. In Figures \ref{figure4}e and f, we utilize the magnetic configuration at simulation time = 5 to depict this X-point and the surrounding field lines. Here, the peripheral field lines of wFR (colored in red) undergo reconnection with the large-scale external toroidal field lines, resulting in the formation of field lines along the small arcades and another set of external field lines. The reconnection scenario described above is highly likely to be the cause of the filament deflection. During this reconnection process, the magnetic pressure decreases at the X-point, thereby attracting wFR to deflect toward this location. This deflection, in turn, further enhances the reconnection rate. As a result, the reconnection and deflection form a positive feedback, ultimately driving wFR to erupt in a non-radial direction. 

\begin{figure*}[ht]
\centering
\includegraphics[scale=0.5]{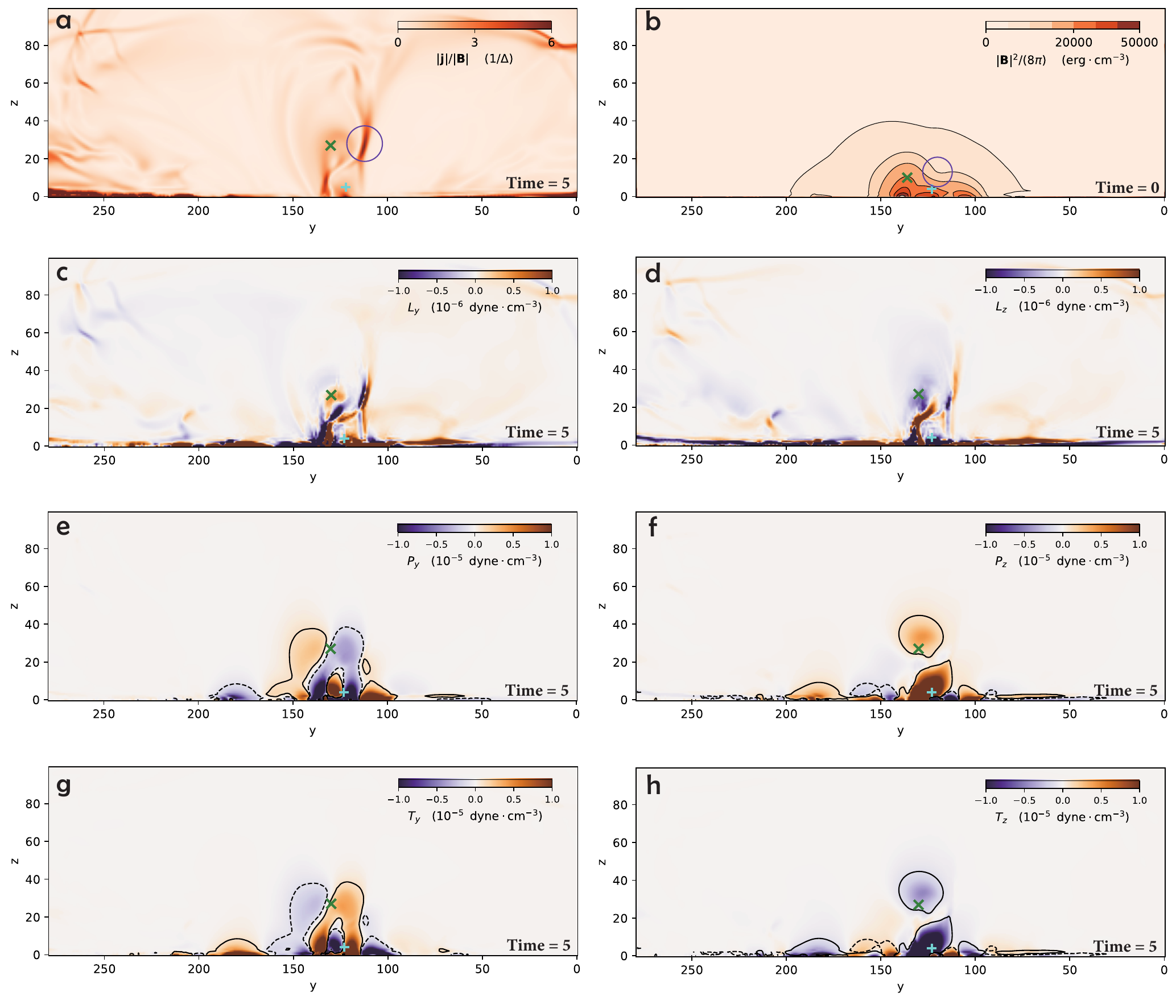}
\caption{(a) The distribution of $|\boldsymbol{j}|/|\boldsymbol{B}|$ on a slice at $x=30.7$ Mm. The purple circle marks the position of current sheet at the X-point. (b) Magnetic energy density at simulation time = 0. The purple circle marks the region of lower magnetic energy density near the flux rope. (c) Lorentz force in the $y$ direction. (d) Lorentz force in the $z$ direction. (e) Magnetic pressure in the $y$ direction. (f) Magnetic pressure in the $z$ direction. (g) Magnetic tension force in the $y$ direction. (h) Magnetic tension force in the $z$ direction. All panels show physical quantity distribution on the same slice with Figures \ref{figure4}a and b. Except panel (b), all panels show quantity distribution at simulation time = 5. The green cross marks the location of wFR's axis, while the cyan plus sign indicates the position of the small arcade. Note that the $y$-axis is reversed to facilitate comparison with Figure \ref{figure4}.}
\label{figure5}
\end{figure*}

To quantitatively analyze the cause of the deflection, we select the same vertical slice as in Figures \ref{figure4}a and b to illustrate the distributions of magnetic energy density and Lorentz force in Figure \ref{figure5}. First, we calculate the ratio of $|\boldsymbol{j}|/|\boldsymbol{B}|$, which clearly identifies the locations of different structures, including wFR, the small arcades, and the current sheet near the X-point (Figure \ref{figure5}a). Next, we compute the magnetic energy density $|\boldsymbol{B}^2|/2\mu_0$. As shown by the contours in Figure \ref{figure5}b, a region of low magnetic energy density is present near wFR, which exerts an attractive force, causing wFR to move toward it. Finally, in Figures \ref{figure5}c, d, e, f, g, and h, we present the distribution of the Lorentz force and its components, including both magnetic pressure force and magnetic tension force, in the $y$ and $z$ directions. Figures \ref{figure5}c and d illustrate the total Lorentz force in the horizontal and vertical directions. Driven by the local minima in magnetic pressure (or equivalently, magnetic energy density), the field lines on both sides of the current sheet, including the peripheral field lines of wFR around the X-point, exhibit convergent motion toward the central current sheet. The vertical Lorentz force in Figure \ref{figure5}d exhibits a contrast between wFR and the small arcades: the lower portion of the wFR experiences a strong upward Lorentz force that drives its ascent, while the entire small arcade system is subjected to a dominant downward Lorentz force that effectively suppresses its eruptive potential. Figures \ref{figure5}e and h provide a direct understanding of the X-point reconnection from the perspective of magnetic pressure and tension force. For the right part of wFR, it experiences magnetic pressure along the negative $y$-axis (Figure \ref{figure5}e), which corresponds to the reduced magnetic pressure caused by X-point magnetic reconnection. Simultaneously, the small arcade is subjected to a vertically downward magnetic tension force, also a consequence of this reconnection, as the newly formed field lines suppress the upward motion of the small arcade. Figure \ref{figure5}g illustrates that, consistent with the twist characteristics of a flux rope, the field lines of wFR experience a magnetic tension force directed toward its axis. Figure \ref{figure5}f demonstrates that both current channels are subjected to an upward magnetic pressure force due to the strong photospheric magnetic field. 

\begin{figure*}[ht]
\centering
\includegraphics[scale=0.5]{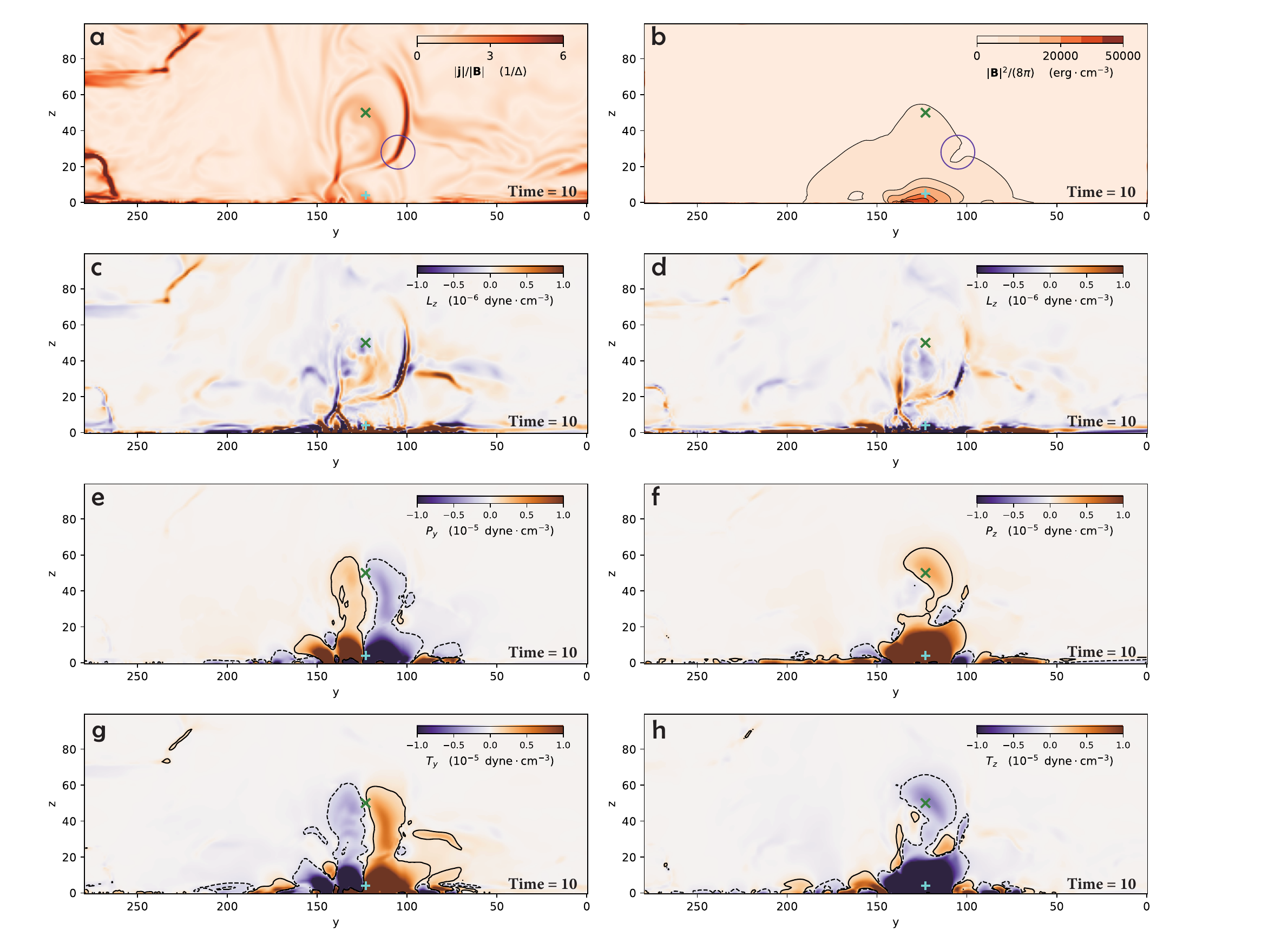}
\caption{Same as Figure~\ref{figure5}, but at simulation time = 10.}
\label{figure6}
\end{figure*}

In order to demonstrate the consistency of the above Lorentz force analysis at later simulation times, when the flux-rope deflection becomes more pronounced, Figure~\ref{figure6} presents the distributions of $|\boldsymbol{j}|/|\boldsymbol{B}|$, magnetic energy density, the Lorentz force, and its individual components at simulation time = 10. The results lead to conclusions similar to those shown in Figure~\ref{figure5}.

\section{Summary and Discussion}\label{discussion}

Through a data-constrained simulation, we successfully reproduce the filament eruption event associated with an X4.5-class solar flare that occurred at 06:18 UT on 2024 May 6. A detailed comparison between simulation results and observational data yields the following key findings:

\begin{enumerate}

\item{The active region exhibited a complex magnetic topology containing two distinct flux rope structures, whose footpoints are spatially correlated with the four hook-shaped features observed in the flare ribbons. The wFR is identified as the source region of the erupting filament. Additionally, a small system of magnetic arcades with opposite helicity are found adjacent to the wFR.}

\item{The deflection mechanism was attributed to the presence of a magnetic X-point configuration formed above two current channels with opposite helicity. Magnetic reconnection occurring in this X-point region resulted in localized magnetic pressure reduction, causing the wFR to deflect toward the null point under the influence of Lorentz forces. Meanwhile, the magnetic tension force of newly formed field lines impeded the eruption of the small arcade below the X-point.}

\end{enumerate}

Besides, the filament eruption displayed pronounced counterclockwise rotation and significant trajectory deflection. The rotational motion is consistent with the negative helicity of the flux rope. In our simulation, the initial twist of the wFR is approximately -0.4, whose absolute value is well below the critical threshold , $\approx 1.25$, required to trigger kink incstability in a line-tied flux rope \citep{Kruskal1954,Hood&Priest1981}. This indicates that kink instability is unlikely to be the primary driver of the observed rotational motion. Instead, the presence of a strong shear component suggests that the filament rotation is more plausibly governed by the influence of the external toroidal magnetic field \citep{Kliem2012, Zhang2025}. As demonstrated in our previous study \citep{Zhang2025}, the external toroidal field can facilitate flux-rope rotation through two mechanisms: (1) promoting the conversion of magnetic twist into axis writhe, and (2) enhancing the lateral Lorentz force acting on the flux rope. Therefore, we conclude that the observed rotation of the filament is predominantly driven by the external magnetic field configuration rather than by internal kink instability.

\subsection{Two competitive current channels}

Prior to the major flare onset, two localized brightenings were observed to the south of wFR. Subsequent analysis reveal these brightenings were interconnected by a small system of magnetic arcades. The magnetic arcades exhibited near-parallel alignment with the wFR while demonstrating opposite magnetic helicity. Both the arcades and wFR originated from the same negative magnetic polarity region, with the arcades positioned above the southern polarity inversion line (sPIL) and the wFR located above the northern PIL (nPIL).

Our numerical simulations demonstrate competitive interaction between current channels forming above these parallel PILs during their formation and evolution. The system exhibits two key characteristics:

\begin{enumerate}

\item{Eruptions preferentially initiate above compact polarity inversion lines (PILs). Our observations demonstrate that flux ropes located above PILs with stronger magnetic field convergence (characterized by steeper magnetic field gradients) tend to erupt first. Quantitative analysis of HMI magnetograms reveals that the magnetic field gradient at the nPIL significantly exceeds that of the sPIL, as evidenced by the nPIL's more compact spatial configuration. These findings align with the numerical simulations of \citet{Bian2022}, who similarly concluded that stronger PILs can accumulate greater non-potential magnetic energy and facilitate the formation of more intense current sheets, thereby enabling more rapid and substantial magnetic energy release during eruptions.}

\item{The rise of wFR suppressed the rise of the arcades. Enhanced convergence motion at the nPIL facilitated more rapid flux cancellation and consequently faster magnetic energy accumulation. This resulted in earlier destabilization and rise initiation of wFR along nPIL compared to the arcades along sPIL. The ascending flux rope was subsequently guided toward the X-point by ongoing reconnection processes, where it further enhanced reconnection rates. This positive feedback mechanism continuously suppressed the rise of the arcades above sPIL.}

\end{enumerate}

\subsection{The "sandwich" magnetic configuration}

Observations of parallel current channels with opposite helicity are not rare. In addition to our findings, \citet{Wang2014,Joshi2016} and \citet{Jing2024} also analyzed analogous configurations in distinct eruptive events. \citet{Wang2014}  identified two parallel loop systems with opposite twists during successive flares (2012 July 6), accompanied by three surges/jets. \citet{Joshi2016} reported interacting filament channels of opposite chirality (2014 April 18–20), triggering three episodes of field-aligned material transport. \citet{Jing2024} proposed a dual-flux-rope interaction model with opposite helicity to explain the initiation mechanism of the M1.5 flare (2014 August 1), as illustrated in their schematic map. These cases collectively demonstrate that helicity reversal in coupled current channels is a recurrent feature in solar active regions, often associated with multiple eruptions and complex magnetic reconnection.

These studies (including ours) reveal a recurrent magnetic configuration characterized by three elongated magnetic polarities arranged in either "+-+" or "-+-" polarity patterns. This configuration produces two parallel PILs, each potentially hosting current channels that manifest as either visible or invisible arcades/flux ropes with opposing helicity signatures. The system's three-dimensional topology features: (1) fishbone-like arcades originating from the central pole and connecting to the bilateral poles; (2) a high-Q region positioned above the fishbone spine, serving as a preferential site for magnetic reconnection; (3) higher-altitude field lines that either connect to remote poles or transition to open field configurations. Following \citet{Wang2014}, we classify this distinctive topology as a "sandwich" magnetic configuration.

The observed magnetic configuration exhibits similarities to classical fan-spine topology \citep{Parnell1996}, but with significant modifications induced by strong shear components. These shear-driven alterations transform the characteristic circular ribbon and bright kernel into three parallel flare ribbons interconnected by a semi-elliptical ribbon, which is similar to the complex flare ribbons in our work. Furthermore, the null point typically present in fan-spine configurations is replaced by a high-Q region in this "sandwich" configuration. The "sandwich" photospheric magnetic configuration likely forms through collision and shear motions between two pairs of sunspots \citep{Chintzoglou2019}. Comparative analysis of previous studies and our work reveals characteristic eruption patterns in this configuration: (1) eruptions preferentially initiate at the more compact PIL \citep{Bian2022}, (2) three distinct parallel flare ribbons are produced with separation motion most pronounced in ribbons flanking the compact PIL. Different eruption dynamics (filament ejection, surge/jet, or material transport) are found in different events, which may be influenced by both the amount of released free energy during the eruption and the connectivity of higher-altitude magnetic field lines. We will investigate this relationship further in future work.

\begin{acknowledgements}

X.M.Z., J.H.G., Y.G., and M.D.D. were supported by the National Key R\&D Program of China (2022YFF0503004, 2021YFA1600504, and 2020YFC2201201) and NSFC (12333009 and 12503063). J.H.G was supported by China National Postdoctoral Program for Innovative Talents fellowship under Grant Number BX20240159. The numerical computation was conducted in the High Performance Computing Center (HPCC) at Nanjing University.

\end{acknowledgements}

\bibliographystyle{aa}
\bibliography{ref}

\begin{thebibliography}{50}
\expandafter\ifx\csname natexlab\endcsname\relax\def\natexlab#1{#1}\fi

\bibitem[{{Bian} {et~al.}(2022){Bian}, {Jiang}, {Feng}, {Zuo}, {Wang}, \& {Wang}}]{Bian2022}
{Bian}, X., {Jiang}, C., {Feng}, X., {et~al.} 2022, \aap, 658, A174

\bibitem[{{Brueckner} {et~al.}(1995){Brueckner}, {Howard}, {Koomen}, {Korendyke}, {Michels}, {Moses}, {Socker}, {Dere}, {Lamy}, {Llebaria}, {Bout}, {Schwenn}, {Simnett}, {Bedford}, \& {Eyles}}]{Brueckner1995}
{Brueckner}, G.~E., {Howard}, R.~A., {Koomen}, M.~J., {et~al.} 1995, \solphys, 162, 357

\bibitem[{{C{\'e}cere} {et~al.}(2023){C{\'e}cere}, {Costa}, {Cremades}, \& {Stenborg}}]{Cecere2023}
{C{\'e}cere}, M., {Costa}, A., {Cremades}, H., \& {Stenborg}, G. 2023, Frontiers in Astronomy and Space Sciences, 10, 1260432

\bibitem[{{Chintzoglou} {et~al.}(2019){Chintzoglou}, {Zhang}, {Cheung}, \& {Kazachenko}}]{Chintzoglou2019}
{Chintzoglou}, G., {Zhang}, J., {Cheung}, M. C.~M., \& {Kazachenko}, M. 2019, \apj, 871, 67

\bibitem[{{Domingo} {et~al.}(1995){Domingo}, {Fleck}, \& {Poland}}]{Domingo1995}
{Domingo}, V., {Fleck}, B., \& {Poland}, A.~I. 1995, \solphys, 162, 1

\bibitem[{{Dud{\'\i}k} {et~al.}(2025){Dud{\'\i}k}, {Aulanier}, {L{\"o}rin{\v{c}}{\'\i}k}, \& {Zemanov{\'a}}}]{Aulanier2025}
{Dud{\'\i}k}, J., {Aulanier}, G., {L{\"o}rin{\v{c}}{\'\i}k}, J., \& {Zemanov{\'a}}, A. 2025, \solphys, 300, 139

\bibitem[{{Gui} {et~al.}(2011){Gui}, {Shen}, {Wang}, {Ye}, {Liu}, {Wang}, \& {Zhao}}]{Gui2011}
{Gui}, B., {Shen}, C., {Wang}, Y., {et~al.} 2011, \solphys, 271, 111

\bibitem[{{Guo} {et~al.}(2023){Guo}, {Qiu}, {Ni}, {Guo}, {Li}, {Gao}, {Schmieder}, {Poedts}, \& {Chen}}]{Guo2023}
{Guo}, J.~H., {Qiu}, Y., {Ni}, Y.~W., {et~al.} 2023, \apj, 956, 119

\bibitem[{{Guo} {et~al.}(2016{\natexlab{a}}){Guo}, {Xia}, \& {Keppens}}]{Guo2016a}
{Guo}, Y., {Xia}, C., \& {Keppens}, R. 2016{\natexlab{a}}, \apj, 828, 83

\bibitem[{{Guo} {et~al.}(2019){Guo}, {Xia}, {Keppens}, {Ding}, \& {Chen}}]{Guo2019}
{Guo}, Y., {Xia}, C., {Keppens}, R., {Ding}, M.~D., \& {Chen}, P.~F. 2019, \apjl, 870, L21

\bibitem[{{Guo} {et~al.}(2016{\natexlab{b}}){Guo}, {Xia}, {Keppens}, \& {Valori}}]{Guo2016b}
{Guo}, Y., {Xia}, C., {Keppens}, R., \& {Valori}, G. 2016{\natexlab{b}}, \apj, 828, 82

\bibitem[{{Guo} {et~al.}(2021){Guo}, {Zhong}, {Ding}, {Chen}, {Xia}, \& {Keppens}}]{Guoy2021}
{Guo}, Y., {Zhong}, Z., {Ding}, M.~D., {et~al.} 2021, \apj, 919, 39

\bibitem[{{Hood} \& {Priest}(1981)}]{Hood&Priest1981}
{Hood}, A.~W. \& {Priest}, E.~R. 1981, Geophysical and Astrophysical Fluid Dynamics, 17, 297

\bibitem[{{Isenberg} \& {Forbes}(2007)}]{Isnberg2007}
{Isenberg}, P.~A. \& {Forbes}, T.~G. 2007, \apj, 670, 1453

\bibitem[{{Jiang} {et~al.}(2009){Jiang}, {Yang}, {Zheng}, {Bi}, \& {Yang}}]{Jiang2009}
{Jiang}, Y., {Yang}, J., {Zheng}, R., {Bi}, Y., \& {Yang}, X. 2009, \apj, 693, 1851

\bibitem[{{Jing} {et~al.}(2024){Jing}, {Lee}, {Mancuso}, {Li}, {Liu}, {Inoue}, {Xu}, \& {Wang}}]{Jing2024}
{Jing}, J., {Lee}, J., {Mancuso}, M., {et~al.} 2024, \apj, 972, 110

\bibitem[{{Joshi} {et~al.}(2016){Joshi}, {Filippov}, {Schmieder}, {Magara}, {Moon}, \& {Uddin}}]{Joshi2016}
{Joshi}, N.~C., {Filippov}, B., {Schmieder}, B., {et~al.} 2016, \apj, 825, 123

\bibitem[{{Kay} {et~al.}(2017){Kay}, {Gopalswamy}, {Xie}, \& {Yashiro}}]{Kay2017}
{Kay}, C., {Gopalswamy}, N., {Xie}, H., \& {Yashiro}, S. 2017, \solphys, 292, 78

\bibitem[{{Kay} {et~al.}(2015){Kay}, {Opher}, \& {Evans}}]{Kay2015}
{Kay}, C., {Opher}, M., \& {Evans}, R.~M. 2015, \apj, 805, 168

\bibitem[{{Keppens} {et~al.}(2023){Keppens}, {Popescu Braileanu}, {Zhou}, {Ruan}, {Xia}, {Guo}, {Claes}, \& {Bacchini}}]{AMRVAC2023}
{Keppens}, R., {Popescu Braileanu}, B., {Zhou}, Y., {et~al.} 2023, \aap, 673, A66

\bibitem[{{Kliem} {et~al.}(2012){Kliem}, {T{\"o}r{\"o}k}, \& {Thompson}}]{Kliem2012}
{Kliem}, B., {T{\"o}r{\"o}k}, T., \& {Thompson}, W.~T. 2012, \solphys, 281, 137

\bibitem[{{Kruskal} \& {Schwarzschild}(1954)}]{Kruskal1954}
{Kruskal}, M. \& {Schwarzschild}, M. 1954, Proceedings of the Royal Society of London Series A, 223, 348

\bibitem[{{Lemen} {et~al.}(2012){Lemen}, {Title}, {Akin}, {Boerner}, {Chou}, {Drake}, {Duncan}, {Edwards}, {Friedlaender}, {Heyman}, {Hurlburt}, {Katz}, {Kushner}, {Levay}, {Lindgren}, {Mathur}, {McFeaters}, {Mitchell}, {Rehse}, {Schrijver}, {Springer}, {Stern}, {Tarbell}, {Wuelser}, {Wolfson}, {Yanari}, {Bookbinder}, {Cheimets}, {Caldwell}, {Deluca}, {Gates}, {Golub}, {Park}, {Podgorski}, {Bush}, {Scherrer}, {Gummin}, {Smith}, {Auker}, {Jerram}, {Pool}, {Soufli}, {Windt}, {Beardsley}, {Clapp}, {Lang}, \& {Waltham}}]{Lemen2012}
{Lemen}, J.~R., {Title}, A.~M., {Akin}, D.~J., {et~al.} 2012, \solphys, 275, 17

\bibitem[{{Liu} {et~al.}(2018){Liu}, {Liu}, {Hu}, {Wang}, \& {Zhao}}]{Liu2018}
{Liu}, Y.~A., {Liu}, Y.~D., {Hu}, H., {Wang}, R., \& {Zhao}, X. 2018, \apj, 854, 126

\bibitem[{{Lugaz} {et~al.}(2012){Lugaz}, {Farrugia}, {Davies}, {M{\"o}stl}, {Davis}, {Roussev}, \& {Temmer}}]{Lugaz2012}
{Lugaz}, N., {Farrugia}, C.~J., {Davies}, J.~A., {et~al.} 2012, \apj, 759, 68

\bibitem[{{Manchester} {et~al.}(2017){Manchester}, {Kilpua}, {Liu}, {Lugaz}, {Riley}, {T{\"o}r{\"o}k}, \& {Vr{\v{s}}nak}}]{Manchester2017}
{Manchester}, W., {Kilpua}, E. K.~J., {Liu}, Y.~D., {et~al.} 2017, \ssr, 212, 1159

\bibitem[{{M{\"u}ller} {et~al.}(2020){M{\"u}ller}, {St. Cyr}, {Zouganelis}, {Gilbert}, {Marsden}, {Nieves-Chinchilla}, {Antonucci}, {Auch{\`e}re}, {Berghmans}, {Horbury}, {Howard}, {Krucker}, {Maksimovic}, {Owen}, {Rochus}, {Rodriguez-Pacheco}, {Romoli}, {Solanki}, {Bruno}, {Carlsson}, {Fludra}, {Harra}, {Hassler}, {Livi}, {Louarn}, {Peter}, {Sch{\"u}hle}, {Teriaca}, {del Toro Iniesta}, {Wimmer-Schweingruber}, {Marsch}, {Velli}, {De Groof}, {Walsh}, \& {Williams}}]{solo}
{M{\"u}ller}, D., {St. Cyr}, O.~C., {Zouganelis}, I., {et~al.} 2020, \aap, 642, A1

\bibitem[{{Pariat} \& {D{\'e}moulin}(2012)}]{Pariat&Démoulin2012}
{Pariat}, E. \& {D{\'e}moulin}, P. 2012, \aap, 541, A78

\bibitem[{{Parnell} {et~al.}(1996){Parnell}, {Smith}, {Neukirch}, \& {Priest}}]{Parnell1996}
{Parnell}, C.~E., {Smith}, J.~M., {Neukirch}, T., \& {Priest}, E.~R. 1996, Physics of Plasmas, 3, 759

\bibitem[{{Pesnell} {et~al.}(2012){Pesnell}, {Thompson}, \& {Chamberlin}}]{Pesnell2012}
{Pesnell}, W.~D., {Thompson}, B.~J., \& {Chamberlin}, P.~C. 2012, \solphys, 275, 3

\bibitem[{{Qiu} {et~al.}(2025){Qiu}, {Guo}, {Ding}, {Li}, {Kong}, \& {Li}}]{Qiu2025}
{Qiu}, Y., {Guo}, Y., {Ding}, M., {et~al.} 2025, \apj, 991, 184

\bibitem[{{Sahade} {et~al.}(2025){Sahade}, {Vourlidas}, \& {Mac Cormack}}]{Sahade2025}
{Sahade}, A., {Vourlidas}, A., \& {Mac Cormack}, C. 2025, \apj, 978, 41

\bibitem[{{Scherrer} {et~al.}(2012){Scherrer}, {Schou}, {Bush}, {Kosovichev}, {Bogart}, {Hoeksema}, {Liu}, {Duvall}, {Zhao}, {Title}, {Schrijver}, {Tarbell}, \& {Tomczyk}}]{Scherrer2012}
{Scherrer}, P.~H., {Schou}, J., {Bush}, R.~I., {et~al.} 2012, \solphys, 275, 207

\bibitem[{{Schou} {et~al.}(2012){Schou}, {Scherrer}, {Bush}, {Wachter}, {Couvidat}, {Rabello-Soares}, {Bogart}, {Hoeksema}, {Liu}, {Duvall}, {Akin}, {Allard}, {Miles}, {Rairden}, {Shine}, {Tarbell}, {Title}, {Wolfson}, {Elmore}, {Norton}, \& {Tomczyk}}]{Schou2012}
{Schou}, J., {Scherrer}, P.~H., {Bush}, R.~I., {et~al.} 2012, \solphys, 275, 229

\bibitem[{{Scott} {et~al.}(2017){Scott}, {Pontin}, \& {Hornig}}]{Scott2017}
{Scott}, R.~B., {Pontin}, D.~I., \& {Hornig}, G. 2017, \apj, 848, 117

\bibitem[{{Shen} {et~al.}(2011){Shen}, {Wang}, {Gui}, {Ye}, \& {Wang}}]{Shen2011}
{Shen}, C., {Wang}, Y., {Gui}, B., {Ye}, P., \& {Wang}, S. 2011, \solphys, 269, 389

\bibitem[{{Titov}(2007)}]{Titov2007}
{Titov}, V.~S. 2007, \apj, 660, 863

\bibitem[{{T{\"o}r{\"o}k} \& {Kliem}(2005)}]{Torok2005}
{T{\"o}r{\"o}k}, T. \& {Kliem}, B. 2005, \apjl, 630, L97

\bibitem[{{Vemareddy} \& {Wiegelmann}(2014)}]{Vemareddy2014}
{Vemareddy}, P. \& {Wiegelmann}, T. 2014, \apj, 792, 40

\bibitem[{{Wang} {et~al.}(2014{\natexlab{a}}){Wang}, {Liu}, {Deng}, {Zeng}, {Xu}, {Jing}, \& {Cao}}]{Wang2014}
{Wang}, H., {Liu}, C., {Deng}, N., {et~al.} 2014{\natexlab{a}}, \apjl, 781, L23

\bibitem[{{Wang} {et~al.}(2014{\natexlab{b}}){Wang}, {Wang}, {Shen}, {Shen}, \& {Lugaz}}]{Wangym2014}
{Wang}, Y., {Wang}, B., {Shen}, C., {Shen}, F., \& {Lugaz}, N. 2014{\natexlab{b}}, Journal of Geophysical Research (Space Physics), 119, 5117

\bibitem[{{Williams} {et~al.}(2005){Williams}, {T{\"o}r{\"o}k}, {D{\'e}moulin}, {van Driel-Gesztelyi}, \& {Kliem}}]{Williams2005}
{Williams}, D.~R., {T{\"o}r{\"o}k}, T., {D{\'e}moulin}, P., {van Driel-Gesztelyi}, L., \& {Kliem}, B. 2005, \apjl, 628, L163

\bibitem[{{Xia} {et~al.}(2018){Xia}, {Teunissen}, {El Mellah}, {Chan{\'e}}, \& {Keppens}}]{Xia2018}
{Xia}, C., {Teunissen}, J., {El Mellah}, I., {Chan{\'e}}, E., \& {Keppens}, R. 2018, \apjs, 234, 30

\bibitem[{{Yang} {et~al.}(2023){Yang}, {Yan}, {Xue}, {Wang}, {Yang}, {Li}, {Xu}, {Peng}, {Sun}, \& {Zhang}}]{Yang2023}
{Yang}, L., {Yan}, X., {Xue}, Z., {et~al.} 2023, \apj, 943, 62

\bibitem[{{Zhang} {et~al.}(2022){Zhang}, {Chen}, {Liu}, \& {Wang}}]{Zhang2022}
{Zhang}, P., {Chen}, J., {Liu}, R., \& {Wang}, C. 2022, \apj, 937, 26

\bibitem[{{Zhang} {et~al.}(2025){Zhang}, {Guo}, {Guo}, \& {Ding}}]{Zhang2025}
{Zhang}, X., {Guo}, J., {Guo}, Y., \& {Ding}, M. 2025, \aap, 695, A1

\bibitem[{{Zhang} {et~al.}(2024){Zhang}, {Guo}, {Guo}, {Ding}, \& {Keppens}}]{Zhang2024}
{Zhang}, X.~M., {Guo}, J.~H., {Guo}, Y., {Ding}, M.~D., \& {Keppens}, R. 2024, \apj, 961, 145

\bibitem[{{Zhao} {et~al.}(2016){Zhao}, {Gilchrist}, {Aulanier}, {Schmieder}, {Pariat}, \& {Li}}]{Zhao2016}
{Zhao}, J., {Gilchrist}, S.~A., {Aulanier}, G., {et~al.} 2016, \apj, 823, 62

\bibitem[{{Zhong} {et~al.}(2021){Zhong}, {Guo}, \& {Ding}}]{Zhong2021}
{Zhong}, Z., {Guo}, Y., \& {Ding}, M.~D. 2021, Nature Communications, 12, 2734

\bibitem[{{Zou} {et~al.}(2019){Zou}, {Jiang}, {Feng}, {Zuo}, {Wang}, \& {Wei}}]{Zou2019}
{Zou}, P., {Jiang}, C., {Feng}, X., {et~al.} 2019, \apj, 870, 97

\end{thebibliography}

\end{document}